\documentclass[
 reprint,
 amsmath,amssymb,
 aps,
]{revtex4-2}
\usepackage{nicefrac,xfrac}
\usepackage{graphicx}
\usepackage{dcolumn}
\usepackage{bm}
\usepackage{xcolor}
\usepackage{multirow}

\usepackage{makecell}
\usepackage{float}
\usepackage{amsmath}
\usepackage{color, colortbl}
\usepackage{soul}
\usepackage{graphicx}
\usepackage{subcaption}
\usepackage{parskip}
\usepackage{SIunits}
\usepackage[colorlinks,bookmarks=false,linkcolor=blue,urlcolor=blue]{hyperref}
\usepackage{enumitem}
\usepackage{dcolumn}
\newcommand{\Hquad}{\hspace{0.75em}} 
\newcommand{\Hquadd}{\hspace{0.5em}} 

\begin{document}

\preprint{APS/123-QED}

\title{Interplay between  ferroelectricity and metallicity in hexagonal YMnO\textsubscript{3}}

\author{Tara N. Tošić}
 \email{tara.tosic@mat.ethz.ch}
 \affiliation{Materials Theory, ETH Zürich, Wolfgang-Pauli-Strasse 27, 8093 Zurich, Switzerland}
 
\author{Yuting Chen}%
\affiliation{Materials Theory, ETH Zürich, Wolfgang-Pauli-Strasse 27, 8093 Zurich, Switzerland}

\author{Nicola A. Spaldin}
\email{nspaldin@ethz.ch}
\affiliation{Materials Theory, ETH Zürich, Wolfgang-Pauli-Strasse 27, 8093 Zurich, Switzerland}

\date{\today}

\begin{abstract}

We use first-principles density functional theory to investigate how the polar distortion is affected by doping in multiferroic hexagonal yttrium manganite, h-YMnO$_3$. While the introduction of charge carriers tends to suppress the polar distortion in conventional ferroelectrics, the behavior in improper geometric ferroelectrics, of which h-YMnO$_3$ is the prototype, has not been studied to date. Using both background charge doping and atomic substitution, we find an approximately linear dependence of the polar distortion on doping concentration, with hole doping suppressing and electron doping enhancing it. We show that this behavior is a direct consequence of the improper geometric nature of the ferroelectricity. In addition to its doping effect, atomic substitution can further suppress or enhance the polar distortion through changes in the local chemistry and geometry.

\end{abstract}

\maketitle


\section{Introduction}

The combination of ferroelectricity and metallicity is seemingly incompatible, because metallic charge carriers tend to screen the long-range interactions responsible for ferroelectricity and inhibit polarization switching by an electric field. In spite of this apparent contra-indication, there have been a number of recent studies on introducing charge carriers into ferroelectric (FE) materials, motivated in part by the possible relevance of polar metals for phenomena such as superconductivity and topology \cite{rishcau2017ferroelectric}. In a device context, materials with polarization robust to charge doping, could be used for example as electrodes to stabilize thin film polarization \cite{puggioni2018polar}. The most widely studied is the prototypical FE BaTiO\textsubscript{3} \cite{Zhou2020review,takahashi2017polar,li2021strain,ma2018strain,raghavan2016probing,michel2021interplay}, where electron doping directly modifies the electronic structure of the formally
$d^0$ Ti$^{4+}$ site, hindering its off-centering through the second-order Jahn-Teller effect and ultimately suppressing the primary FE order parameter \cite{michel2021interplay}.\\

Geometric FEs are promising candidates for persistence of the polar distortion on doping \cite{hickox2023polar,bhowalspaldin2022polar}, since their FE distortion is driven by a structural instability such as a polyhedral rotation that is not caused by the second-order Jahn-Teller effect. As a result, their polarization is expected to be less sensitive to the detailed electronic occupation of the atomic orbitals. Note that, since ferroelectricity is not well defined for a metal, we will refer to the FE distortion as a polar distortion in doped samples. There have been a number of studies on the effect of doping in {\it proper} geometric FEs, in which the polar geometric mode is the primary order parameter connecting the paraelectric (PE) and FE structures. For example, in the A\textsubscript{n}B\textsubscript{n}O\textsubscript{3n+2} Carpy-Galy family, a structural distortion consisting largely of rotations of the oxygen octahedra gives rise to the polarisation. First-principles computations find the polarisation in La\textsubscript{2}Ti\textsubscript{2}O\textsubscript{7} and Sr\textsubscript{2}Nb\textsubscript{2}O\textsubscript{7} to be robust to doping \cite{zhao2018meta}, consistent with the observed coexistence of non-centrosymmetricity and quasi 1D conductivity  \cite{lichtenberg2001synthesis}. On the other hand, ferroelectricity in \textit{improper} geometric FEs emerges as a secondary effect, through coupling to a non-polar structural order parameter which drives the FE phase transition. To our knowledge, the only studies on the coexistence of itinerant charges and ferroelectricity in improper FEs have been done on the Ruddlesden-Popper phase Ca\textsubscript{3}Ti\textsubscript{2}O\textsubscript{7}. In this material, the polar mode couples to two tilt-modes through a trilinear coupling term \cite{benedek2011hybrid}. Here, optical second-harmonic imaging and atomic-resolution scanning transmission electron microscopy have measured formation of polar domains coexisting with metallicity \cite{lei2018observation,levanyuk1974improper}, which was further confirmed by density functional theory \cite{zhang2023optical, huang2019major}.\\

\begin{figure}[h!]
    \centering
    \includegraphics[width=1\linewidth]{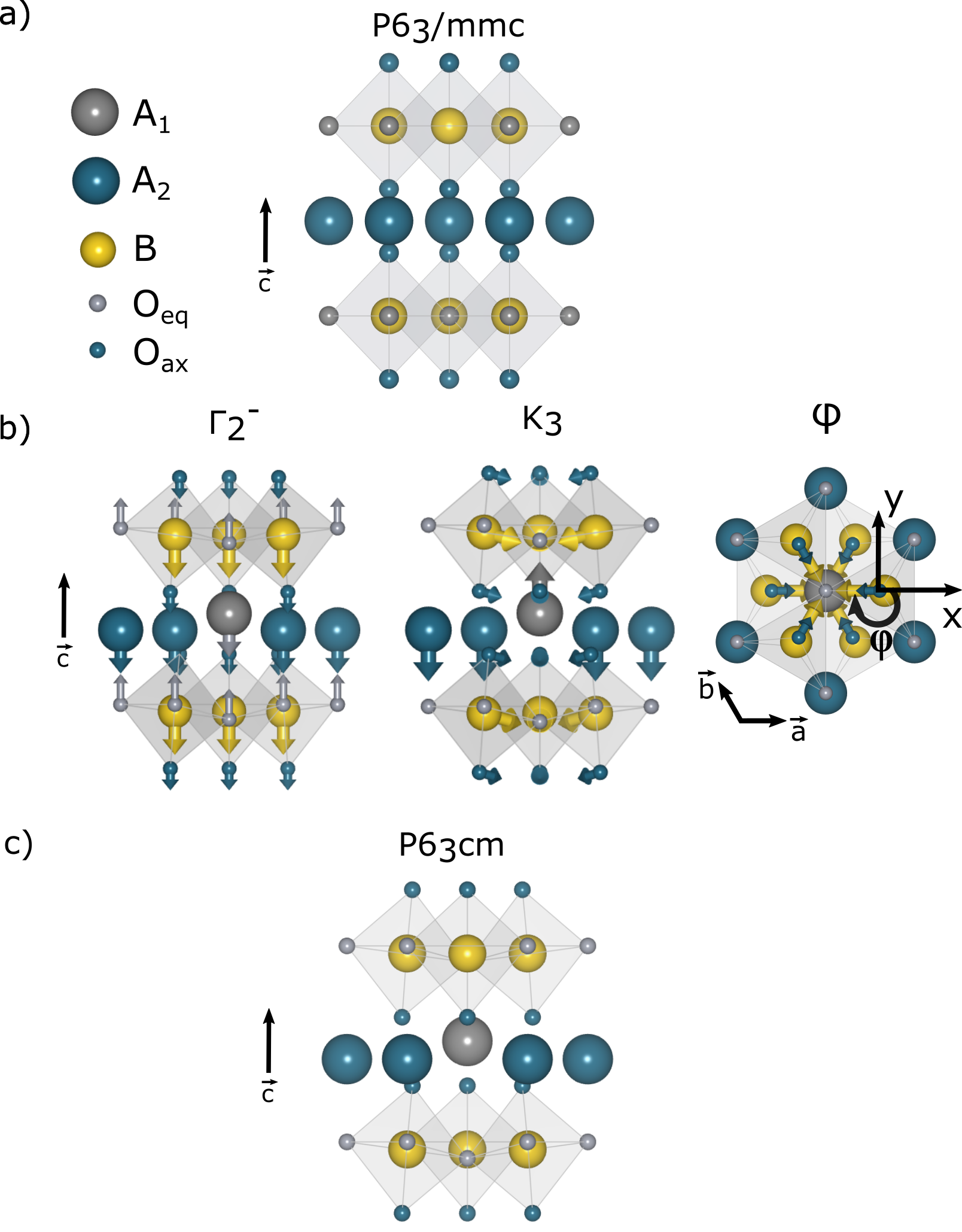}
    \caption{a) Structure of h-YMnO\textsubscript{3} in the PE P6\textsubscript{3}/mmc structure b) The two modes involved in the FEphase transition. Left: the polar $\Gamma_2^-$ mode which creates a polar distortion through relative displacements (shown with arrows) of the ions. Middle and right: the primary FE zone-boundary $K_3$ mode which involves tilting of the MnO\textsubscript{5} bipyramids towards their corner sharing O sites and a buckling of the A\textsuperscript{-} ion planes. The angle $\phi$ characterizes the phase of the bipyramidal tilts. c) Structure of h-YMnO\textsubscript{3} in the FE P6\textsubscript{3}cm structure.}
    \label{structure_modes}
\end{figure}

Motivated by the promising studies on doped hybrid geometric FEs, this work investigates the effects of doping on the prototypical geometric improper FE, h-YMnO\textsubscript{3}. h-YMnO\textsubscript{3} belongs to the family of hexagonal manganites h-\textit{R}MnO\textsubscript{3}, where  \textit{R} denotes In, Y, Sc  or small radius rare-earth elements (Dy-Lu)) \cite{artyukhin2014landau,meier2017global,skjaervo2019unconventional,meier2020manifestation,nordlander2019ultrathin}. The structure is composed of planes of corner-shared MnO\textsubscript{5} trigonal bipyramids (Mn\textsuperscript{3+} and O\textsuperscript{2-}), alternating with triangular layers of Y\textsuperscript{3+} ions. At high temperature, the structure is centrosymmetric with the P6$_3$/mmc space group, (see Fig.~\ref{structure_modes}a)) \cite{lukaszewicz1974x,gibbs2011hightemp}. Below $T_c\simeq 1270$\,K, the MnO\textsubscript{5} trigonal bipyramids tilt towards trimerization centers, inducing a buckling of each Y layer \cite{fennie2005ferroelectric, meier2017global,skjaervo2019unconventional} as shown in the right panel of Fig.~\ref{structure_modes}b). This structural distortion is associated with the zone-boundary $K_3$ mode of the high-symmetry structure, which triples the unit cell and lowers the symmetry to the polar P6$_3$cm space group (shown in Fig.~\ref{structure_modes}c)), but with no polarization. We write its associated order parameter as $\mathbf{Q_{\text{K$_3$}}}=Q_{\text{K$_3$}}(\cos{(\phi}),\sin{(\phi)})$, where $Q_{\text{K$_3$}}$ describes the mode's amplitude and $\phi=n\frac{\pi}{3}$ its phase ($n$ is an integer), as illustrated in Fig.~\ref{structure_modes}b). A secondary zone center $\Gamma_2^-$ structural distortion in which anions and cations shift in opposite directions along $c$ couples to K\textsubscript{3}, inducing a net polarization described by the scalar order parameter $Q_{\Gamma_2^-}$. Note that this two-mode coupling distinguishes this material from the previously studied hybrid improper Ca$_3$Ti$_2$O$_7$, where a three-mode coupling term stabilizes the polar displacements \cite{lei2018observation}. The two modes couple non-linearly and the Landau free energy in terms of $\mathbf{Q_{\text{K$_3$}}}$ and $Q_{\Gamma_2^-}$ up to fourth order is \cite{fennie2005ferroelectric,artyukhin2014landau}:

\begin{align}\label{landauPQ}
   E(\mathbf{Q_{\text{K$_3$}}},Q_{\Gamma_2^-}) &= \alpha Q_{\text{K$_3$}}^2 + \beta Q_{\text{K$_3$}}^4+ \alpha_P Q_{\Gamma_2^-}^2 + \beta_P Q_{\Gamma_2^-}^4 \nonumber \\
   &+ \gamma Q_{\text{K$_3$}}^3Q_{\Gamma_2^-}\cos{(3\phi)} +\gamma' Q_{\text{K$_3$}}^2Q_{\Gamma_2^-}^2\quad ,
\end{align}

where the direction of the K\textsubscript{3} displacement, and hence the sign of the polarization, are set by the value of $\phi$ (see
Fig.1b)). Since the polar distortion arises as a secondary effect through coupling to the K\textsubscript{3} primary order \cite{artyukhin2014landau,fennie2005ferroelectric,meier2017global}, we anticipate that, compared to conventional FEs, it will again, as in the hybrid improper compound, be robust to the presence of dopant charges, leading to a possible large region of coexistence between the polar distortion and metallicity.\\ 

\section{Computational details}

\subsection{DFT details}

Our calculations are performed within the density functional theory (DFT) framework using the planewave pseudopotential VASP code \cite{dronskowski1993crystal,kresse1994ab,kresse1996efficiency,kresse1996efficient}. We use the projector-augmented wave pseudopotentials \cite{blochl1994projector,PhysRevB.59.1758} Y\_sv, Mn\_sv and O from the VASP library with \textit{(4s)\textsuperscript{2}(4p)\textsuperscript{6}(5s)\textsuperscript{2}(4d)\textsuperscript{1}}, \textit{(3s)\textsuperscript{2}(4s)\textsuperscript{1}(3p)\textsuperscript{6}(3d)\textsuperscript{6}} and \textit{(2s)\textsuperscript{2}(2p)\textsuperscript{4}} valence configurations, respectively. For the substitution atoms, we use the Mg\_sv, Ca\_sv, Hf, Ti\_sv and V\_sv projector-augmented wave pseudopotentials, with the following valence shell electrons: \textit{(2s)\textsuperscript{2}(3s)\textsuperscript{2}(2p)\textsuperscript{6}}, \textit{(3s)\textsuperscript{2}(4s)\textsuperscript{2}(3p)\textsuperscript{6}}, \textit{(5d)\textsuperscript{3}\,(6s)\textsuperscript{1}}, \textit{(3s)\textsuperscript{2}(4s)\textsuperscript{1}(3p)\textsuperscript{6}(3d)\textsuperscript{3}} and \textit{(3s)\textsuperscript{2}(4s)\textsuperscript{1}(3p)\textsuperscript{6}(3d)\textsuperscript{4}}. We treat the exchange correlation functional using the local density approximation (LDA) + Hubbard U approach \cite{perdew1981self}, with U=5.2\,eV on Mn $d$ orbitals and 6.5\,eV on V $d$ orbitals, calculated using the linear response Ansatz \cite{coccioni2005linear} (see Supplementary Information). We use a $\Gamma$-centered k-point grid of $6\times6\times3$ for the full structural relaxation and further relax the internal coordinates with a $12\times12\times6$ k-grid upon introduction of the charges. Densities of states (DOS) are calculated using an $18\times18\times9$ grid. We perform non-constrained collinear magnetic calculations, with magnetic moments on the Mn ions (initialized to 4\textmu B) parallel to each other within each layer and anti-ferromagnetically aligned between the consecutive layers, to preserve the structural symmetry.\\

\subsection{Charge doping}

In the first part of our work, charge doping is simulated by changing the total number of valence electrons in the system and adding a compensating background charge to maintain charge neutrality by modifying the NELECT flag in VASP. Since the relaxation of the lattice parameters is not well-defined during this process \cite{bruneval2015pressure}, we fix the lattice constants either to their fully relaxed undoped low symmetry values (FE lattice parameters) or fully relaxed high symmetry values (PE lattice parameters).\\

\subsection{Substitution doping}

In the second part of our work, we dope by impurity atom substitution, and single out three factors that influence $P$: the size of the impurity atom and the subsequent structural changes, the role of the charge carrier, and the variations in lattice constants. We quantify their separate contributions as follows: starting from a relaxed h-YMnO\textsubscript{3} cell, we first introduce the impurity atom but subtract the added charges. For example, in the case of the quadrivalent Hf dopant substituted on the A\textsubscript{1} or A\textsubscript{2} site, we subtract one electron and add a background charge to maintain charge neutrality. In a second step, we relax the structure containing the dopant atom and do not modify the charge. The lattice constants are fixed in the first two steps. Finally, we remove the constraint on the lattice parameters and let the whole cell relax. By comparing the polar distortion of the undoped cell with that calculated after each step, we can quantify the different contributions to $P$ stemming from the size of the impurity atom, the role of the charge carrier and the change in lattice constants, respectively. We name the aforementioned three separate contributions the \textit{impurity atom effect}, the \textit{charge carrier effect} and the \textit{lattice constant effect}.

\subsection{Calculation of polar distortion and COHP}

Since we are interested in doped systems where the Berry phase polarization is undefined, we calculate the size of the polar distortion, $P$, by summing the displacements ($d_i$) of the ions from their reference positions in the high-symmetry structure, multiplied by their Born Effective Charges (BECs), $Z^*_i$:

\begin{equation}\label{eq:BEC_P}
P = \frac{e}{\Omega}\sum_iZ_i^*d_i\quad ,
\end{equation}

where $e$ is the elementary electronic charge and $\Omega$ the unit cell volume.\\

We use the following BEC values calculated for the FE structure in reference \cite{van2004origin}: $Z_{\text{Y}}^*=+3.6$, $Z_{\text{Mn}}^*=+3.3$, $Z_{\text{O\textsubscript{ax}}^*}=-2.3$ and $Z_{\text{O\textsubscript{eq}}^*}=-2.2$ (where O\textsubscript{ax} and O\textsubscript{eq} refer to the axial and equatorial oxygens, respectively, see Fig.~\ref{structure_modes}). Note that due to the geometric nature of the ferroelectricity, the BEC values are close to the formal charges \cite{filippetti2003strong}. Using these values, we obtain $P$=8.06\,\textmu C/cm\textsuperscript{2} for our fully relaxed undoped structure. \\ 

Bond strength is characterized using crystal orbital Hamiltonian populations (COHPs) and integrated crystal orbital Hamiltonian populations (ICOHPs) calculated using the LOBSTER package \cite{maintz2013analytic,maintz2016lobster,nelson2020lobster}, with an integration energy range between $\rm E_f-10$\,eV and $\rm E_f$, capturing the occupied Mn $3d$ and O $2p$ states. Our -ICOHP plots indicate the change in bonding strength as a function of doping. The input WAVECAR file for the COHP calculations is obtained in VASP without symmetrization of the charge density (ISYM flag is set to -1) and a k-point grid of $12\times12\times6$.\\


\section{Results and discussion}
We begin by presenting the main results of our calculations on the effects of background charge doping on the polar distortion.

\subsection{Charge doping}

\begin{figure}
    \centering
    \includegraphics[width=0.9\linewidth]{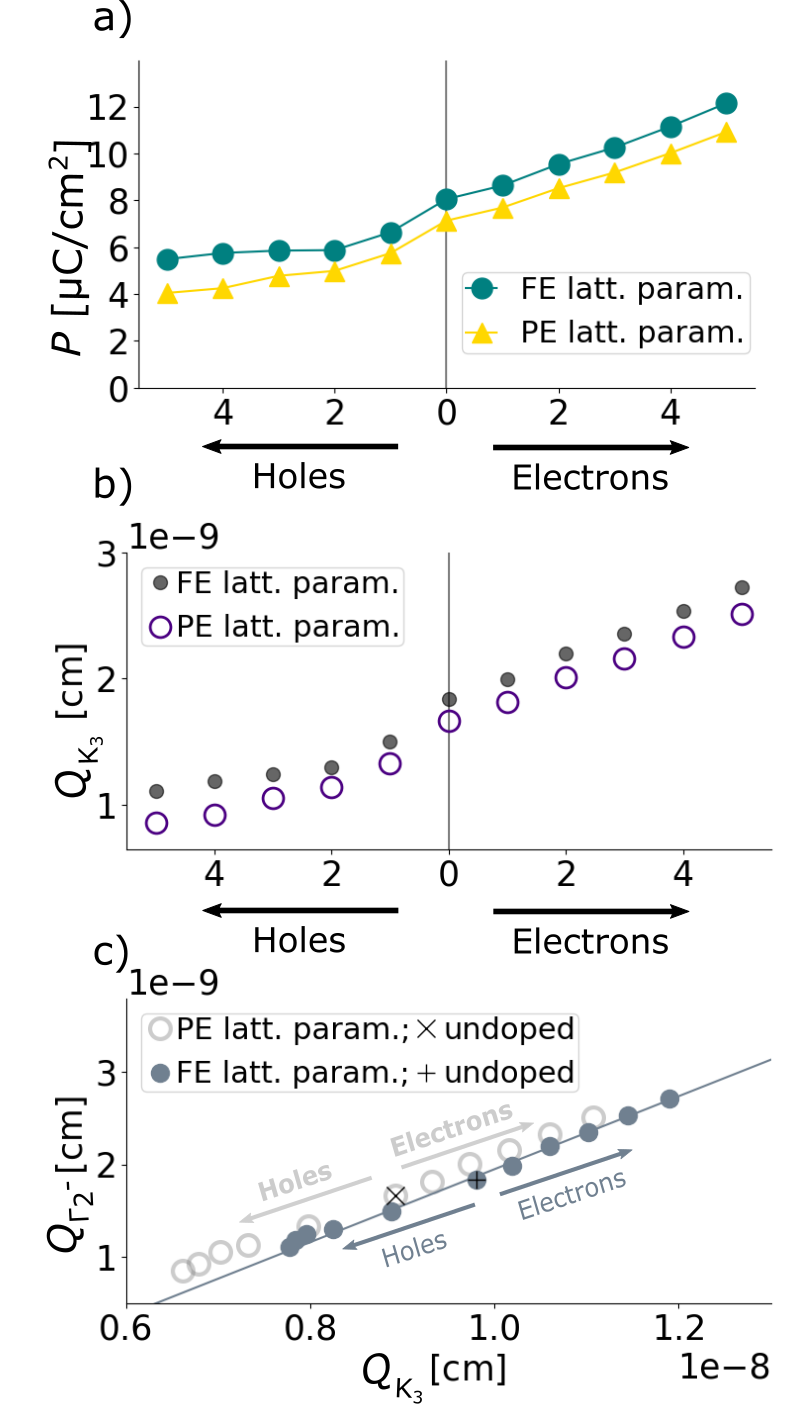}
    \caption{a) Polar distortion as a function of hole and electron doping concentration at the lattice parameters of the fully relaxed FE and PE structures. b) Amplitude ($Q_{\text{K$_3$}}$) of the primary FE distortion mode K\textsubscript{3} as a function of doping concentration at FE and PE lattice parameters. c) $Q_{\Gamma_2^-}$ plotted as a function of $Q_{\text{K$_3$}}$ evaluated for the range of hole and electron doping, for FE and PE lattice parameters. The gray line shows $P\simeq 0.4Q_{\text{K$_3$}}$, which we use as the reference trend for the rest of our analysis.}
    \label{fig:P_vs_dope}
\end{figure}

We begin by calculating the structure as a function of background-charge electron and hole doping concentration at fixed PE and FE lattice parameters and extracting the size of the polar distortion following Eq. \ref{eq:BEC_P}. We use our calculated undoped lattice constants which are slightly underestimated for both the FE P6\textsubscript{3}cm ($a=6.05$ \AA, $c=11.32$ \AA) and PE P6\textsubscript{3}/mmc ($a=3.53$ \AA, $c=11.13$ \AA) structures compared to measured values ($a=6.14$ \AA, $c=11.41$ \AA ~and $a=3.61$ \AA, $c=11.39$ \AA, respectively) \cite{vanaken2001hexagonal,lukaszewicz1974x} as well as to the DFT-calculated values in Ref. \cite{fennie2005ferroelectric}, which used an LSDA pseudopotential and U=8\,eV. We show the size of our calculated polar distortion as a function of electron and hole doping concentration in Fig. \ref{fig:P_vs_dope}a). In contrast to the conventional FE BaTiO\textsubscript{3}, where both charge carriers suppress $P$, in this case $P$ increases monotonically with electron doping: the added electrons do not hinder the ferroelectricity and even promote it. Hole doping, on the other hand, lowers $P$ but does not fully suppress it. We also note that, while the polar distortion remains slightly larger for the FE lattice parameters than for the PE lattice parameters at all doping values, both curves follow the same trend. \\

\begin{figure*}
\begin{minipage}{1\linewidth}
  \includegraphics[width=1\linewidth]{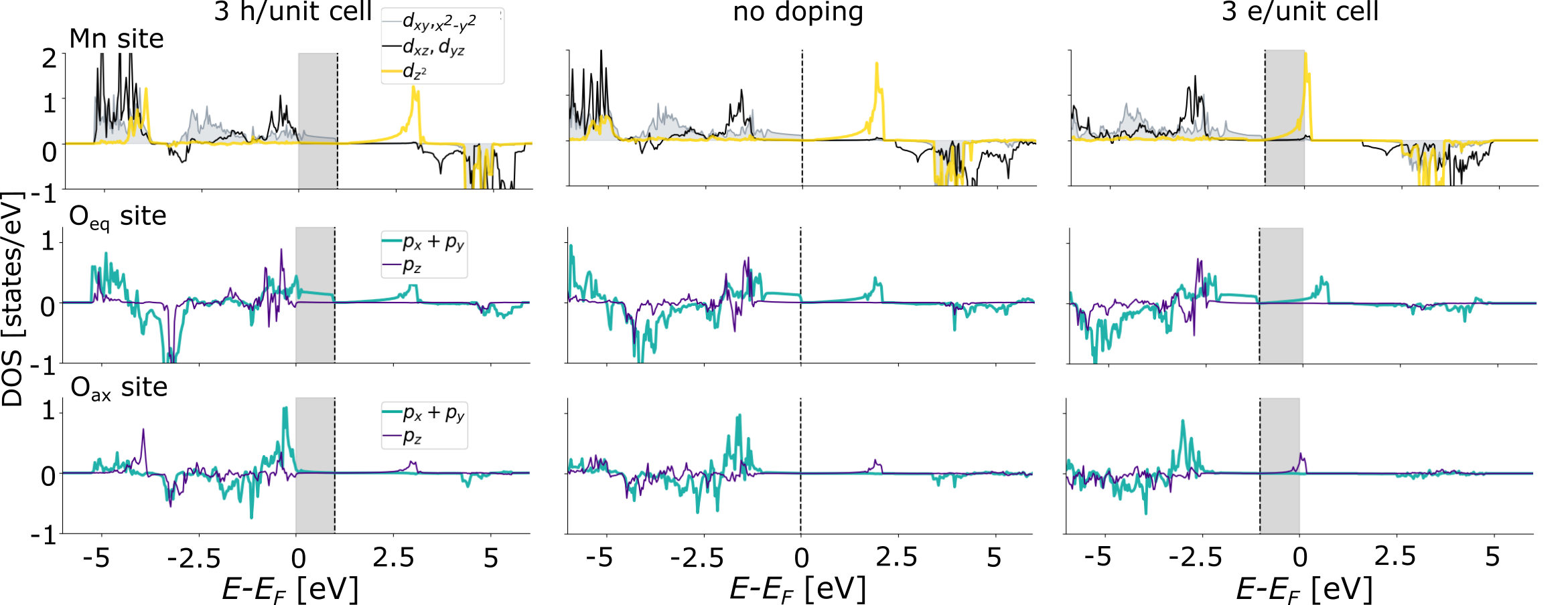}
\end{minipage}
\begin{minipage}{0.9\linewidth}
  \includegraphics[width=0.95\linewidth]{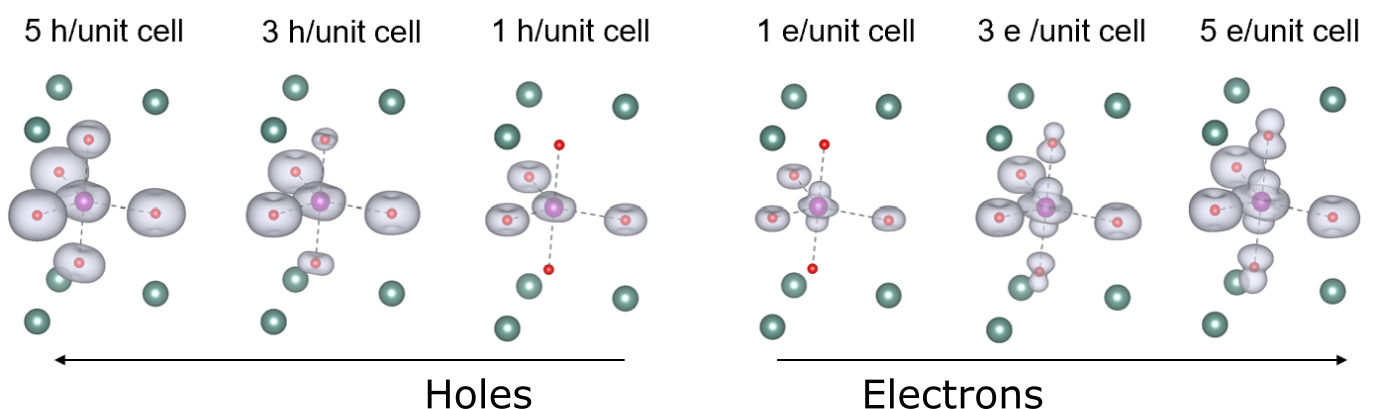}
  \centering
\end{minipage}
\caption{Upper panel: orbital-resolved density of states for (left) 3 dopant holes per 30 atom unit cell, (middle) no doping and (right) 3 dopant electrons per 30 atom unit cell (left, middle and right columns, respectively) in h-YMnO\textsubscript{3} for the Mn, O\textsubscript{eq} and O\textsubscript{ax} sites (top, middle and bottom rows, respectively). Vertical dotted line indicates the top of the valence band in the undoped compound, with the shaded gray areas indicating the change in occupied states on doping. Lower panel: change in charge density in the conduction band when doped with electrons (right) and in the valence band and when doped with holes (left). The isosurface is plotted at 0.005 e/bohr\textsuperscript{3} in both cases. }
  \label{fig:charge_redistribution}
\end{figure*}

 To uncover the cause of this change in $P$, we first analyze the structural changes which occur upon doping by performing a symmetry mode analysis using the AMPLIMODES program \cite{orobengoa2009amplimodes,perez2010mode}. Fig. \ref{fig:P_vs_dope}b) shows the dependence of the primary FE phase transition order parameter, $Q_{\text{K$_3$}}$, on charge doping, for FE and PE lattice parameters. We see that the evolution of $Q_{\text{K$_3$}}$ as a function of charge is similar to that of $Q_{\Gamma_2^-}$ in Fig.~\ref{fig:P_vs_dope}a) for both sets of lattice parameters. We confirm this in Fig.~\ref{fig:P_vs_dope}c) where we plot $Q_{\Gamma_2^-}$ versus $Q_{\text{K$_3$}}$ and find a linear dependence, consistent with Landau theory at large $Q_{\text{K$_3$}}$ \cite{fennie2005ferroelectric}. In the following, we investigate the electronic causes for these changes in $Q_{\text{K$_3$}}$ and $Q_{\Gamma_2^-}$ upon introduction of charges.\\

\begin{figure}
  \centering
    \includegraphics[width=0.8\linewidth]{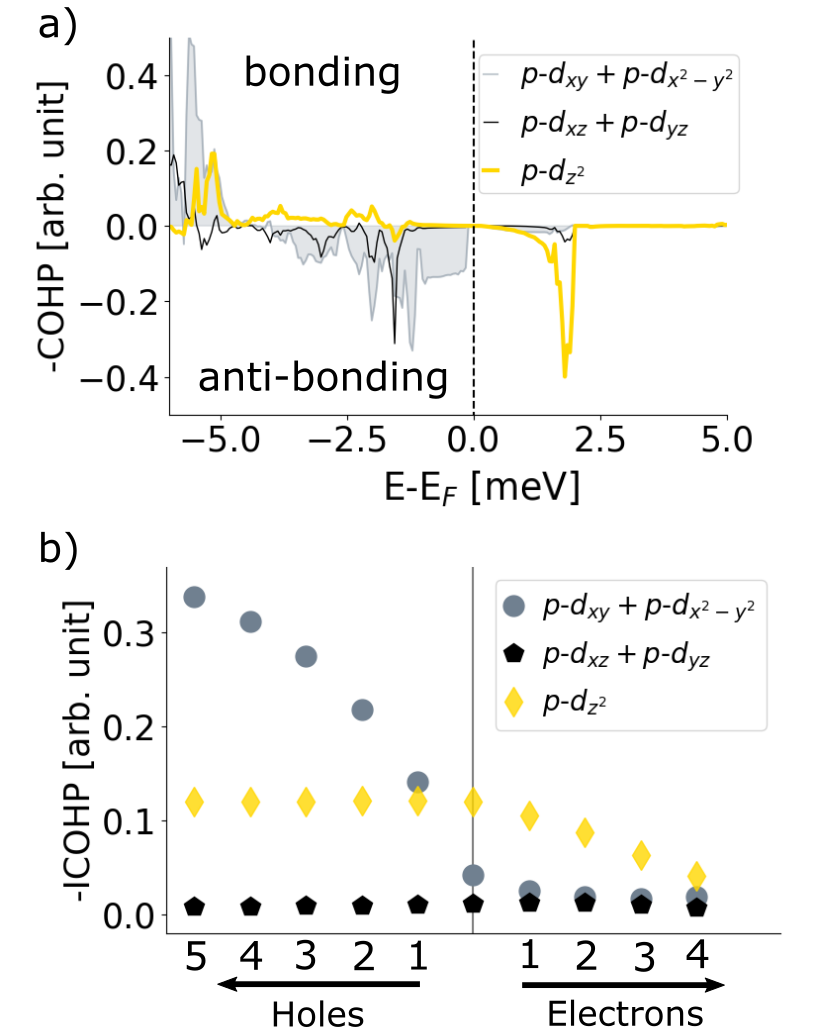}
    \caption{a) -COHP between O\textsubscript{eq} $p$ orbitals and Mn $d_{xy}/d_{x^2-y^2}$, Mn $d_{xz}/d_{yz}$ and Mn $d_{z^2}$ orbitals, for undoped h-YMnO\textsubscript{3}. Positive and negative values correspond to bonding and anti-bonding states, respectively. b) -ICOHP between pairs of O\textsubscript{eq} $p$ and Mn $d$ orbitals as a function of doping.}
    \label{fig:icohp_cohp}
\end{figure}

To understand the increase and decrease of $Q_{\text{K$_3$}}$, $Q_{\Gamma_2^-}$ and hence $P$, with electron and hole doping, respectively, we next calculate the electronic DOS and the charge density as a function of doping, and show our results in Fig.\,\ref{fig:charge_redistribution}. In the upper panel, we plot our calculated atom- and orbital-resolved DOSs for undoped h-YMnO\textsubscript{3}, as well as for representative electron and hole concentrations of three charge carriers per 30 atom unit cell. The black-dotted vertical line indicates the top of the valence band in the undoped compound and the gray shaded rectangle indicates the shift in $E_F$ upon doping, with the edge without the black-dotted vertical line indicating the $E_F$ value in the doped compounds. When no charges are introduced, the valence band states closest to the top of the valence band are the hybridized Mn $d_{xy},d_{x^2-y^2}$ and O\textsubscript{eq} $p_x, p_y$ orbitals, while the Mn $d_{xz}$, d$_{yz}$ states hybridize with the O\textsubscript{eq} $p_z$ and O\textsubscript{ax} $p_x,p_y$ states at 1 to 1.5 \,eV below the valence bond edge. Our calculated band is gap is equal to $\simeq 1.3$\,meV and the conduction band consists of Mn $d_{z^2}$, O\textsubscript{eq} $p_x,p_y$ and the O\textsubscript{ax} $p_z$ orbitals. In the case of electron doping (right panel of Fig.\,\ref{fig:charge_redistribution}), $E_f$ moves into the conduction band, with electrons first occupying the Mn $d_{z^2}$ states and then, at 1 electron per unit cell, the O\textsubscript{eq} $p_x+p_y$ orbital; this is clearly seen in the charge density, shown below the DOS in Fig.\,\ref{fig:charge_redistribution}, and in the occupied DOS Mn $d_{z^2}$ and O\textsubscript{eq} $p_x+p_y$ tails in Fig.~S3. As electron doping increases, the occupation of these states increases further and the dopant electrons also start occupying the O\textsubscript{ax} $p_z$ orbitals (see the 3 and 5 electron/unit cell DOS and charge density plots in Fig.~\ref{fig:charge_redistribution} and Fig.~S3). In the case of hole doping, $\rm E_f$ moves into the valence band (Fig.~\ref{fig:charge_redistribution} left panel) and the depletion of electrons is initially most pronounced from the Mn $d_{x^2-y^2}$, $d_{xy}$ and O\textsubscript{eq} $p_x+p_y$ orbitals. As hole doping increases, Mn $d_z^2$ and O\textsubscript{ax} $p$ states also start depleting. We see in the charge density plots below the DOSs that the charge state of the Y remains effectively unchanged.\\

Two hybridizations are allowed by symmetry between the planar O\textsubscript{eq} $p$ orbitals and the occupied Mn $d$ orbitals in the ideal MnO\textsubscript{5} bipyramidal environment: $p-d_{xz}/d_{yz}$ 
 and $p-d_{xy}/d_{x^2-y^2}$ (note that, as a result of the MnO\textsubscript{5} tilt in the P6\textsubscript{3}cm structure, small $p_z-d_{xy}/d_{x^2-y^2}$ and $p_{x,y}-d_{z^2}$ hybridizations are also allowed, consistent with optical spectroscopy measurements  \cite{choi2008electronic}). To determine how the hybridization evolves with doping, we evaluate the Crystal Orbital Hamilton Population (COHP) for the undoped compound (shown in Fig.~\ref{fig:icohp_cohp}a)) as well as the Integrated Crystal Orbital Hamilton Population (ICOHP) up to E\textsubscript{F} as a function of doping concentration (shown in Fig.~\ref{fig:icohp_cohp}b)). \\

In the undoped COHP in Fig.~\ref{fig:icohp_cohp}a), the band edges are formed of anti-bonding O\,$p-$Mn\,$d$ states. This indicates that the electron doping will increase O\,$p-$Mn\,$d$ $ab$ character and weaken Mn-O bonds, whereas hole doping will decrease O\,$p-$Mn\,$d$ $ab$ character and strengthen Mn-O bonds. In detail, we find the $p-d_z^2$ interactions at the bottom of the conduction band, and $p-d_{xy}+p-d_{x^2-y^2}$ at the top of the valence band. Anti-bonding $p-d_{xz}+p-d_{yz}$ interactions occur at $-2.5$\,eV below the top of the valence band.\\

On electron doping, electrons start occupying the $p-d_{z^2}$ dominated yellow anti-bonding states at the bottom of the conduction band. This weakens the planar Mn-O\textsubscript{eq} bond, and is reflected in a reduction in the size of the $p-d_z^2$ ICOHP (yellow diamonds in Fig.~\ref{fig:icohp_cohp}b)). This reduction in in-plane bonding allows for larger $Q_{\text{K$_3$}}$ polyhedral tilting and an in turn increase in $P$.\\
 
On hole doping, the anti-bonding $p-d_{xy}/d_{x^2-y^2}$ bonds, at the top of the valence band, are depleted, increasing the corresponding bond strength, as seen in the increase in the size of the ICOHP (grey circles). This in turn causes a stronger in-plane Mn-O\textsubscript{eq} bond, consistent with a smaller MnO\textsubscript{5} bipyramidal K\textsubscript{3} tilt and a reduced $P$.\\

The $p-d_{xz}/d_{yz}$ interactions contribute almost negligibly to the change in bond strength on either electron or hole doping, as their energy levels are not close to the band edges. The ICOHP is close to zero for the range of background doping we considered, and even at high doping value of 5 holes/unit cell, the Mn $d_{xz}$ and $d_{yz}$ orbitals are only slightly depleted (Fig.~S3).\\

In summary we have shown that dopant electrons increase $P$, while dopant holes suppress it. We additionally calculate that $Q_{\text{K$_3$}}$ increases (decreases) with electron (hole) doping and that this $\Delta Q_{\text{K$_3$}}$ change is a result of the weakening (strengthening) of the in-plane Mn-O bonding. This in turn explains the response in $\Delta P$ through the linear $Q_{\text{K$_3$}}-Q_{\Gamma_2^-}$ coupling.\\


\subsection{Impurity atoms}

\begin{figure*}
  \includegraphics[width=0.9\linewidth]{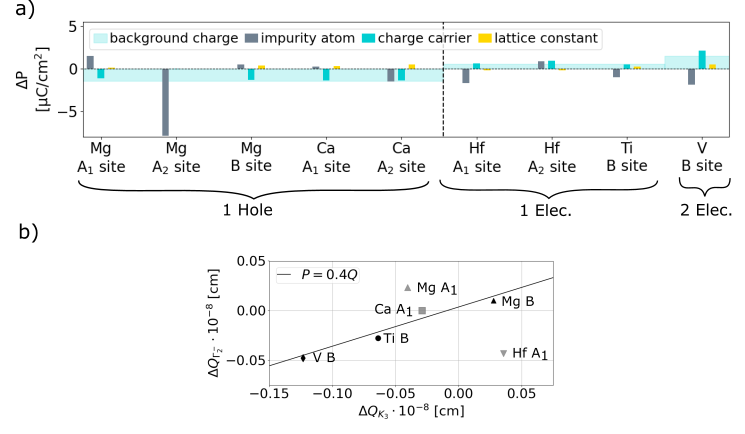}
  \centering
\caption{a) Contribution to $P$ from introduction of charge carriers and changes in lattice constants. Faded turquoise level indicates the change in $P$ due to the corresponding background charge doping concentration shown previously in Fig.~\ref{fig:P_vs_dope}). b) Impurity atom effects in substitution compounds which preserve P6\textsubscript{3}cm symmetry. $Q_{\Gamma_2^-}$ versus $Q_{\text{K$_3$}}$ compared to the linear trend calculated for the background charge method: $Q_{\Gamma_2^-}=0.4Q_{\Gamma_2^-}$.}
  \label{fig:P_contribution}
\end{figure*}

In the second part of the study, we introduce charge carriers by replacing either one Y\textsuperscript{3+} (A\textsubscript{1} or A\textsubscript{2} site atoms) or one Mn\textsuperscript{3+} (B site atom), by one impurity atom per 30 atom unit cell. We choose the following substitution atoms: Ca\textsuperscript{2+} (1 hole per unit cell), Mg\textsuperscript{2+} (1 hole per unit cell), Hf\textsuperscript{4+} (1 electron per unit cell), Ti\textsuperscript{4+} (1 electron per unit cell) and V\textsuperscript{5+} (2 electrons per unit cell).\\

Substitution doping creates more complex changes in the crystal's chemistry and geometry than simple background charge doping. Therefore, in order to disentangle the different contributions to the changes in $P$, we adopt the three step approach introduced in Ref. \cite{michel2021interplay} to separate the contributions from the charge carrier (\textit{charge carrier effect}), lattice constants (\textit{lattice constant effect}) and other factors resulting from the impurity atom (\textit{impurity atom effect}), as described in the Computational Details section. Our results for the charge carrier and lattice constant contributions are shown in Fig.~\ref{fig:P_contribution}a) and we separately show the observed impurity atom effects in Fig.~\ref{fig:P_contribution}b). We summarize our results from this section in Table~\ref{Table:2}, listing the ionic radii and charge carrier contributions and the calculated corresponding lattice constants, polar distortion $P$ as well as the change in polar distortion $\Delta P$ from the unsubstituted case, for all the substitution atoms.\\

First we note that, across all substituted compounds (excluding Mg A\textsubscript{2}-site substitution which is a special case that we will discuss later), the \textit{charge carrier effect} on the polar distortion (turquoise bars in Fig.~\ref{fig:P_contribution}a)) is consistent with our results from the background-charge method (turquoise shading in Fig.~\ref{fig:P_contribution}a)). We see that electron doping tends to promote the polar distortion while hole doping suppresses it, and the sizes of the changes in $P$ from the impurity atom charge carriers closely match those of the background-charge calculations.\\

Secondly, concerning the \textit{lattice constant effect}, we find that changes in lattice constants (yellow bars in Fig.~\ref{fig:P_contribution}a)) contribute negligibly to the change in polar distortion: they increase $P$ by a small amount for most substitution compounds apart from the Hf-substituted compound, where we calculate an associated small decrease in $P$. This correlates with a decrease in $c$ in the Hf-substituted compound, compared to an increase in $c$ for all the other substitution elements,  as shown in Table \ref{Table:2}. Our calculations are consistent with previous X-ray diffraction experiments, correlating an increase (decrease) in $c$ with an increase (decrease) in the $K_3$ mode amplitude which in turn, through its coupling to $\Gamma_2^-$, promotes (suppresses) the FE polar distortion \cite{katsufuji2002crystal}. Note the contrast to conventional FEs such as BaTiO\textsubscript{3}, where lattice contributions are important, and where elongations of the lattice parameter along the polar direction increase $P$ \cite{michel2021interplay}.\\

Finally, we analyze the change in $P$ caused by factors other than the \textit{charge} and \textit{lattice constant effects}, referring to them collectively as \textit{impurity atom effects}. We find that the  A\textsubscript{1}- and B-site substituted compounds preserve the P6\textsubscript{3}cm symmetry of pure YMnO$_3$: we analyze them in terms of the $Q_{\text{K$_3$}}$ and $Q_{\Gamma_2^-}$ ($\propto P$) described above. We quantify the order parameter changes using AMPLIMODES~\cite{orobengoa2009amplimodes}, by replacing the identity of the substitution element in the structure file with the original Y or Mn ion of the host after the DFT relaxations, and present the resulting changes in $Q_{\text{K$_3$}}$ and $Q_{\Gamma_2^-}$ in Fig.~\ref{fig:P_contribution}b).\\

Fig.~\ref{fig:P_contribution}b) shows the change in $Q_{\Gamma_2^-}$ as a function of $Q_{\text{K$_3$}}$, both owing to the \textit{impurity atom effect}. We observe that the $(Q_{\text{K$_3$}},P)$ coupling closely follows the linear relationship predicted by Landau theory. Interestingly, the slope is the same as obtained in our earlier background charge calculations ($P=0.4Q_{\text{K$_3$}}$, plotted here again for comparison). This suggests that the proportionality constant $\gamma$ (see Eq.~\ref{landauPQ}) is characteristic for our system and does not change much when substituting on the B site. We note also that the sizes of the changes in $Q_{\text{K$_3$}}$ arising from B-site impurity atom effects are comparable to those stemming from the charge effect, and are related to the radius of the substituting atom, with $Q_{\text{K$_3$}}$ decreasing (increasing) for B-site substitution atoms with smaller (larger) radii. For A$_1$-site substitution, $Q_{\text{K$_3$}}$ is larger for smaller radius substitution atoms as there is less hindrance to tilting but $Q_{\Gamma_2^-}$ is smaller due to additional shifts of the A-site cations. Our corresponding increases or decreases in $P$ are in agreement with Ref.~\cite{sverre2020domain}  \\

\begin{table*}
\footnotesize
\ldots
\newcolumntype{d}[1]{D{.}{\cdot}{#1} }
\begin{tabular}{ m{2cm} m{1cm} m{2.5cm} m{2.5cm} m{1.4cm} m{1.4cm} m{2cm} m{2cm} m{2cm}}
 \hline
 Dopant & Site & Impurity atom & Charge carrier & \multicolumn{2}{l}{Lattice constants} & $P$  & \Hquad$\Delta P$ & $\Delta P$\\
 & & Ionic radius [Å] &  & $V$ [Å\textsuperscript{3}] & $c$ [Å] & [\textmu C/cm$^2$] & \Hquad[\%] & [\textmu C/cm$^2$]\\
 \hline
 Undoped & - & - & - & 360.4 & 11.33 & 8.06 & \Hquad\Hquadd 0 & \Hquad 0\\
 \hline
 Background & - & -  & 1h & 360.4 & 11.33 & 6.64& \Hquadd$-0.2$ & $-0.02$\\
 
 \multirow{3}*{Mg\textsuperscript{2+}} & A\textsubscript{1} & 0.89  $\downarrow$ & 1h & 349.9 & 11.35 & 8.56 &\Hquadd$+6.3$ & $+0.51$\\
 & A\textsubscript{2}\textsuperscript{$\star$} & 0.89  $\downarrow$ & 1h & 349.1 & 11.38 & 0.08 &$-99.0$ &$-7.98$\\
 & B & 0.89  $\uparrow$& 1h &359.2& 11.42 &7.55 &\Hquadd$-6.3$  & $-0.51$\\
 
 \multirow{2}*{Ca\textsuperscript{2+}} & A\textsubscript{1} & \multirow{2}*{1.12  $\uparrow$} & 1h & 357.9 & 11.42 & 7.20 & $-10.7$ & $-0.86$\\
 & A$_2^{\star}$& & 1h & 357.9 & 11.43 & 5.61 & $-30.4$ &$-2.45$\\
 \hline
 Background & - & -  & 1e & 360.4 & 11.33 & 8.64 & \Hquadd$+0.1$&$+0.01$\\
 \multirow{2}*{Hf\textsuperscript{4+}} & A$_1^\star$ & \multirow{2}*{0.83  $\downarrow$} & 1e & 357.3 & 11.27 & 6.76& $-16.1$&$-1.30$\\
  & A\textsubscript{2} & & 1e & 357.3 & 11.26 & 9.71 & $+20.6$&$+1.66$\\
 Ti\textsuperscript{4+} & B & 0.51 $\downarrow$ & 1e & 363.8 & 11.44 & 7.79 & \Hquadd$-3.3$&$-0.27$\\
 \hline
 Background & - & -  & 2e & 360.4 & 11.33 & 9.55& \Hquadd$+0.2$&$+0.02$\\
 V\textsuperscript{5+} & B & 0.46 $\downarrow$ & 2e & 365.9 & 11.50 & 8.81 & \Hquadd$+9.4$&$+0.76$\\
 \hline
\end{tabular}
\centering
\caption{Summary table of the calculated total changes in the substituted compounds. We list substitution site, Shannon ionic radius \cite{shannon1976revised}, number and type of introduced charge carriers, cell volume, $c$ lattice parameter and $P$ and change in $P$ for all considered doping elements. The radii of Y\textsuperscript{3+} and Mn\textsuperscript{3+} are 1.019~\AA~ and 0.58~\AA, respectively \cite{shannon1976revised}. $\star$ indicates the lowest A-substitution site energy, and $\uparrow$ and $\downarrow$ indicate that the substitution atom's radius is larger or smaller than the substituted atom's radius, respectively.}
\label{Table:2}
\end{table*}

A\textsubscript{2}-site substitution affects the $O_ax-A_2$ bond above the trimerization center, disturbing the buckling of A layer and changing the structural symmetry of the compound. Therefore, we do not analyze the structures in terms of $Q_{\Gamma_2^-}$ and $Q_{\text{K$_3$}}$. Instead, we note that, as the radius of the substitution ion decreases, the substitution atom increasingly rattles with respect to the other A\textsubscript{2} site. $P$ either increases or decreases, depending on the direction of the substitution atom's displacement along $c$. The Ca positive A$_2$ site ion moves down along $c$, giving a $30\%$ suppression of $P$, whereas A$_2$-site Hf moves up resulting in a $20\%$ increase in $P$. Finally, Mg A$_2$ substitution is a special case, with the radius of the Mg$^{2+}$ ion accidentally resulting in the down-middle-up A-layer configuration, characteristic of the PE P$\bar{3}$c symmetry \cite{huang2013delicate,kumagai2012observation,huang2014duality,griffin2017defect}.\\

As a concluding comment, we point out that for A-site impurity substitutions, we find that the substitution site leading to lower $P$ is energetically more favorable in each case than the higher $P$ compound; while Ca substitution energetically favors the A\textsubscript{2} site, Hf prefers to substitute on the A\textsubscript{1} site, and Mg substitution favors the PE A\textsubscript{2}-substituted compound. \\

\section{Conclusions}

In summary, we have calculated and rationalized the relationship between doping and FE polar distortion in the improper FE h-YMnO\textsubscript{3}.\\

Our main finding is that the polar distortion persists in doped h-YMnO\textsubscript{3},  with electron doping increasing $P$, and hole doping reducing it, both linearly proportional to the doping concentration over the entire range studied. This result is in stark contrast to the behavior of conventional FEs,  where polar distortion is suppressed for both hole and electron doping even at small doping concentrations. Our analysis of the changes in chemical bonding on doping indicates that the planar Mn-O bonds weaken (strengthen) upon electron (hole) doping which in turn increases (decreases) the magnitude of the primary order parameter of the FE structure. \\

This main result holds for both background-charge doping and doping via explicit atomic substitution. In the case of atomic substitution, we find additional contributions to the total $P$ from structural changes caused by the changes in lattice parameters or the impurity atom effect itself, which, in the latter case, can dominate over the change in electron or hole count. Our calculations suggest that one should B-site impurity atom substitute with electron charge carriers, such as Ti\textsuperscript{4+} or V\textsuperscript{5+}, in order to enhance $P$.\\

Our findings indicate that doped improper geometric FEs are a promising candidate for new polar metals. In the case of the multiferroic hexagonal manganites, the additional magnetic degree of freedom could lead to an intriguing interplay between metallicity, ferroelectricity and magnetization. We hope that the results that we present here motivate future experimental and theoretical work in this direction.  

\section{Acknowledgments}
We thank Veronica F. Michel and Tobias Esswein for useful discussions regarding the computational details of this study. This work was funded by the European
Research Council (ERC) under the European Union’s
Horizon 2020 research and innovation program project
HERO grant agreement No. 810451. Computational resources were provided by ETH Zürich and the Swiss National Supercomputing center, project IDs eth3 and s1128.

\bibliography{ref.bib}

\end{document}